\documentclass[conference]{IEEEtran}

\usepackage[utf8]{inputenc}
\usepackage[T1]{fontenc}
\usepackage[swedish,english]{babel}

\usepackage{graphicx, float, subfigure, blindtext}

\newcommand\IEEEhyperrefsetup{
bookmarks=true,bookmarksnumbered=true,%
colorlinks=true,linkcolor={black},citecolor={black},urlcolor={black}%
}

\usepackage[\IEEEhyperrefsetup, pdftex]{hyperref}

\usepackage{mathtools}

\usepackage{cite}
\usepackage{amsmath,amssymb,amsfonts}
\usepackage{graphicx}
\usepackage{textcomp}
\usepackage{multirow}
\usepackage{amsmath}
\usepackage{algorithm}
\usepackage[noend]{algpseudocode}
\usepackage{xcolor}

\makeatletter
\algnewcommand{\LineComment}[1]{\Statex \hskip\ALG@thistlm \(\triangleright\) #1}
\def\BState{\State\hskip-\ALG@thistlm}
\makeatother

\definecolor{xanadu}{rgb}{0.45, 0.53, 0.47}

\usepackage{url} 
\def\BibTeX{{\rm B\kern-.05em{\sc i\kern-.025em b}\kern-.08em
    T\kern-.1667em\lower.7ex\hbox{E}\kern-.125emX}}
\newcommand{\centered}[1]{\begin{tabular}[c]{@{}l@{}} #1 \end{tabular}}

\usepackage[nolist,nohyperlinks]{acronym}
\usepackage{cleveref}
\graphicspath{{images/}}

\usepackage{titlesec}
\titlespacing*{\section}{0pt}{*1}{*1}
\titlespacing*{\subsection}{0pt}{*1}{*1}
\renewcommand{\thesubsubsection}{\arabic{subsubsection}}
\titleformat{\subsubsection}[runin]{\itshape}{\thesubsubsection)}{1em}{}[:]
\titlespacing*{\subsubsection}{\parindent}{0pt}{*1}

\author{\IEEEauthorblockN{ %
Alireza Dehlaghi-Ghadim\IEEEauthorrefmark{1}
Mahshid Helali Moghadam\IEEEauthorrefmark{2}, 
Ali Balador\IEEEauthorrefmark{3}
Hans Hansson\IEEEauthorrefmark{4}
}
\IEEEauthorblockA{
School of Innovation, Design and, Engineering\\
Mälardalens University, Västerås, Sweden\\
Research Institute of Sweden (RISE)\\\\
Email:
\IEEEauthorrefmark{1}alireza.dehlaghi.ghadim@(mdu.se/ri.se), 
\IEEEauthorrefmark{2}mahshid.helali.moghadam@mdu.se,\\
\IEEEauthorrefmark{3}li.balador@mdu.se 
\IEEEauthorrefmark{4}hans.hansson@(mdu.se/ri.se)
}}

\title{Anomaly Detection Dataset for Industrial Control Systems}
\begin{document}
\maketitle

\begin{abstract}
Over the past few decades, Industrial Control Systems (ICSs) have been targeted by cyberattacks and are becoming increasingly vulnerable as more ICSs are connected to the internet. Using Machine Learning (ML) for Intrusion Detection Systems (IDSs) is a promising approach for ICS cyber protection, but the lack of suitable datasets for evaluating ML algorithms is a challenge. Although there are a few commonly used datasets, they may not reflect realistic ICS network data, lack necessary features for effective anomaly detection, or be outdated.
This paper presents the \textit{'ICS-Flow'} dataset, which offers network data and process state variables logs for supervised and unsupervised ML-based IDS assessment. The network data includes normal and anomalous network packets and flows captured from simulated ICS components and emulated networks. The anomalies were injected into the system through various attack techniques commonly used by hackers to modify network traffic and compromise ICSs. We also proposed open-source tools, ``ICSFlowGenerator" for generating network flow parameters from Raw network packets. The final dataset comprises over 25,000,000 raw network packets, network flow records, and process variable logs. The paper describes the methodology used to collect and label the dataset and provides a detailed data analysis. Finally, we implement several ML models, including the decision tree, random forest, and artificial neural network to detect anomalies and attacks, demonstrating that our dataset can be used effectively for training intrusion detection ML models.

\end{abstract} 


\section{Introduction}
\label{C:sec:introduction}

Industrial Control Systems (ICSs) are used to control various industrial processes, such as power plants, power grids, railways, and factories \cite{stouffer2011guide}. However, many of the protocols used lack encryption or authentication mechanisms, making them vulnerable to cyberattacks \cite{conti2021survey}. Recent years have seen an increase in cyber threats to industrial systems \cite{cheng2018industrial}, with notable examples including the Stuxnet malware attack on Iranian Uranium enrichment facilities \cite{falliere2011w32}, 
hacking Ukrainian power grid \cite{alladi2020industrial}, 
the Irongate malware attack on Siemence's ICS \cite{kumar2016irongate},
and the Triton malware attack on the Saudi Arabian petrochemical plant \cite{di2018triton}. 
Furthermore, Kaspersky ICS-CERT report \cite{kasperskywebsite} shows that 33.8\% of ICS computers were compromised in 2021, indicating that cybersecurity is a major concern for modern ICSs. This is why ICS security has become a major topic of research in recent years \cite{schwab2018state}, \cite{filkins2019sans}.

To protect industrial systems against the increasing risk of cyberattacks, Intrusion Detection Systems (IDSs) are introduced \cite{sharafaldin2018toward}. These systems monitor networks or hosts to detect cyberattacks or malicious activities \cite{ahmim2018intrusion}, notifying security administrators or event management systems of any detected threats. An effective IDS should detect attacks accurately in a minimum time with the least number of false-positive alerts. To achieve these goals, using Machine Learning (ML) \cite{umer2022machine} and Deep Learning (DL) \cite{luo2021deep} in IDSs seems promising. However, due to security concerns and the risk of interrupting industrial processes, it is not feasible to test intrusion detection methods on operational industrial systems. Alternately, a dataset  containing a range of cyberattacks could provide a benchmark to compare and evaluate intrusion detection algorithms \cite{hamid2018benchmark} before deploying them in real-world scenarios.

Creating a comprehensive and representative dataset of intrusions in ICSs would provide an invaluable resource for the development and testing of new detection methods, as well as for training and validating ML algorithms. It would enable researchers to evaluate the effectiveness of different approaches to intrusion detection, identify areas for improvement, and enhance the overall robustness and effectiveness of cybersecurity measures for ICSs. However, despite its potential benefits, using such datasets can be challenging due to various obstacles. Here, we outline some of these challenges:

\begin{itemize}
    \item Currently, only a limited number of datasets are available to evaluate ML-based anomaly detection in ICSs \cite{gomez2019generation}. However, some of these datasets are based on unrealistic implementations.   
    \item An important consideration for anomaly detection is the nature of the anomaly. For intrusion detection tasks, only a small subset of available datasets are relevant because they inject anomalies through cyberattacks.    
    \item Some ICS testbeds lack crucial details or have implemented them incorrectly, which can impact the accuracy and effectiveness of anomaly detection methods.
    \item Some datasets are highly anonymized and cannot be shared due to confidentiality concerns, while others do not reflect current market trends \cite{sharafaldin2018toward}.
    \item A significant challenge in training ML models for anomaly detection is the lack of labeled data. Some ICS anomaly datasets are unlabeled, making it challenging to train effective models.
    \item Network trace labeling in available datasets is typically based on the automatic generation of synthetic network traces, which eliminate necessary details for accurately distinguishing between legitimate and malicious activity.
    \item Available datasets for anomaly detection tasks vary in the type of data logged, with some only recording process state variables, some only recording controlling commands, and others capturing entire network packets. This study focuses specifically on the latter type of data, which can limit the availability of suitable datasets.
    \item Due to the highly specialized nature of IDSs, transferring datasets between them is difficult, and customized datasets for each domain are preferred \cite{gomez2019generation}.
\end{itemize}

 Moreover, in network-based intrusion detection for the ICS domain, the existing datasets are often unsuitable for training intrusion detection models since they are designed for labeling individual network traces, and most are based on the automatic generation of synthetic network traces \cite{guerra2022datasets}.
Anomaly detection in ICSs differs from anomalous network packet detection because anomalies may exist across entire traffic, not just in a specific network packet. Therefore, there is a need to redefine the intrusion detection task to include anomaly detection in the network traffic pattern rather than just a specific network packet. Therefore, we aim to provide our dataset in both forms of raw packet files and network flow records, which is suitable for studying network flow patterns. 

This article aims to provide an intrusion detection dataset that addresses the mentioned issues by providing a realistic benchmark to compare ML-based IDSs in the ICS domain. To achieve this goal, we leverage the ICSSIM framework \cite{dehlaghi2023icssim}, a tool for simulating customized virtual ICS security testbeds, to investigate cyber threats and attacks. In particular:

\begin{enumerate}
    \item We implement four types of cyberattacks: 'Reconnaissance', 'Distributed Denial of Service' (DDoS), 'false data injection' using Man-in-the-Middle (MitM) technique, and 'Replay' attacks on an ICS testbed. 
    \item  We develop ICSFlowGenerator, a reusable open-source tool to extract network flow features from raw network packets.  
    \item We propose the publicly available 'ICS-Flow' dataset\footnote{\url{https://www.kaggle.com/datasets/alirezadehlaghi/icssim}} as a public resource. This dataset is unique in several ways:
    \begin{enumerate}
        \item To further enhance the dataset's practicality, we have implemented multiple labeling strategies.
        \item It contains a diverse set of network flow features that capture different aspects of ICS network behavior.
        \item The anomalies in the dataset are due to the realistic implementation of network attack scenarios. This differs from other datasets that employ the creation of synthetic network anomalies, making our dataset more representative of real-world conditions.
        \item We offer a complete dataset without any anonymization to provide comprehensive support for anomaly detection, including a network flow dataset, process state snapshots, attack logs, and ICS components logs.
    \end{enumerate}   
\end{enumerate}

Moreover, to demonstrate the practicality of our proposed dataset, we evaluate various ML-based intrusion detection models on the 'ICS-Flow' dataset and compare their performance.

The remainder of this paper is organized as follows. Section \ref{C:sec:background} presents the current state of the art for available testbeds. Section \ref{C:sec:mainwork} proposes the ICS-Flow intrusion detection dataset by implementing a scenario using our proposed testbed in another paper and injecting different types of attacks.
Section \ref{C:sec:datasetmetrics} provides a detailed analysis of the generated dataset.
Section \ref{C:sec:ai}  shows the implementation of anomaly detection techniques on the proposed dataset. Section \ref{C:sec:results} compares and discusses the results of the anomaly detection methods. 
Section \ref{C:sec:validity} identifies the primary threats to the study's validity and applied mitigation techniques.
Finally, Section \ref{C:sec:conclusion} concludes the paper and provides future research directions.

\section{Related Work}
\label{C:sec:background}

This section briefly surveys related work on network intrusion detection datasets to highlight the prerequisites for future datasets. A comprehensive survey of available datasets and their strengths and weaknesses for anomaly detection is beyond the scope of this paper. However, there are several surveys in this area, such as \cite{ghurab2021detailed, ferrag2020deep, ring2019survey}, to mention a few.\\

Many datasets are available for anomaly detection in computer networks, but the KDD Cup '99\footnote{\url{http://kdd.ics.uci.edu/databases/kddcup99/kddcup99.html}} dataset is the most well-known and widely used. In 1998, a DARPA dataset \cite{lippmann2000evaluating} was collected using TCPdump of US air force LAN simulation. A subset of the DARPA dataset used to extract features by the Massachusetts Institute of Technology (MIT) Lincoln Laboratory results in KDD Cup '99 dataset \cite{protic2018review}. This dataset contains 41 features per connection, categorized into three groups: basic features, traffic features, and content features. These features are computed for normal and attack records, belonging to four types of attacks: User to Root (U2R), Remote to Local attack (R2L), Probing, and Denial of Service attack. However, there is some criticism against this dataset. Firstly underlying network traffic backs to a few decades ago. Secondly, the dataset records are not refined well since there are many redundant records, and record classes are not balanced across the dataset \cite{mahoney2003analysis}.\\

In 2009, Mahbod Tavallaee et al. \cite{tavallaee2009detailed} conducted a statistical analysis on the KDDCUP'99, finding that some issues with the dataset adversely affected the anomaly detection experiences. They enhanced the KDD Cup dataset by resolving major criticism such as duplicate or irrelevant records and data asymmetry, which led to a new dataset titled NSL-KDD\footnote{\url{http://www.unb.ca/cic/datasets/nsl.html}}. This dataset collected approximately 150,000 data points and divided them into training and test subsets with the same attributes as KDD-CUP 99. This means that this data set does not represent modern network attack scenarios.

Morris et al. \cite{morris2014industrial} introduced comprehensive datasets encompassing network traffic, process control, and process measurement features gathered from two laboratory-scale ICSs - a gas pipeline and a water storage tank - utilizing the Modbus over serial lines protocol. Their research involved executing various attacks on their experimental setups, such as reconnaissance, response injection, command injection, and Denial-of-Service (DoS) attacks. Nonetheless, it is important to note that their network traffic dataset solely comprises the features of Modbus commands utilized within the ICSs.

To generate a dataset that reflects modern network traffic scenarios and contains varieties of low-footprint attacks, the Australian centre for cybersecurity researchers generated the UNSW-NB 15\footnote{\url{https://research.unsw.edu.au/projects/unsw-nb15-dataset}} dataset \cite{moustafa2015unsw}. This dataset contains 41 features, categorized into six groups: flow, basic, content, time, additional general, and connection features. The first four feature categories were derived directly from dumped traffic, and the auxiliary C\# program was developed to compute the other two additional groups of features. They used IXIA PefectStorm Tool \footnote{\url{http://www.ixiacom.com/products/perfectstorm}} to generate normal and abnormal traffic; therefore, despite the simulation of 9 different attack types, their attack implementation restricts to predefined synthetic attack types in the IXIA tool.

Canadian Institute for cybersecurity proposed the CICIDS2017\footnote{\url{https://www.unb.ca/cic/datasets/ids-2017.html}} dataset based on profiling the human users' abstract behavior  \cite{sharafaldin2018toward}. To generate realistic background traffic, they profiled the behavior of 25 human users, considering eight famous attacks, namely brute force, DoS, botnet, and Heartbleed. Finally, they extracted more than 80 network parameters and then, using Random Forest Regressor, selected the best features for each attack. Although the CICIDS2017 dataset has been used in many intrusion detection experiments, it is not suitable for ICS since it does not reflect the ICS network traffic patterns.


Mathur and Tippenhauer created a small-scale, but fully operational water treatment system (SWaT) testbed for cybersecurity research \cite{mathur2016swat}. Since SWaT is large and geographically dispersed, Programmable Logic Controllers (PLC) use Wireless and wired Fieldbus communications to control the sensors and actuators. Several attack scenarios are defined for this testbed which targets single or multi-points on single or multi-stages. They released the final dataset containing process state variables, selected packet features, and logs of performed attacks \cite{goh2016dataset}\footnote{\url{ https://itrust.sutd.edu.sg/itrust-labs_datasets/dataset_info/}}.  

\begin{table*}[]

\begin{center}
\caption{Summary of Intrusion detection datasets related to our proposed dataset}
\label{C:tab:comparision}
\begin{tabular}{ |p{1.6cm}||p{0.6cm}|p{3.3cm}|p{4.1cm}|p{2.7cm}|  }
 \hline
 \textbf{Dataset}   &   \textbf{Year}   &   \textbf{Data Source}   &    \textbf{Records}  & \textbf{Attacks}
 \\\hline
 
DARPA \cite{lippmann2000evaluating} &
1998 &
\centered{Network traffic and audit logs\\ collected using a simulated\\ network (air force LAN)} &
\centered{ Raw Pcap files of simulated network \\dump } &
\centered{   - U2R \\    - R2L \\    - DoS \\    - Probe \\}

\\\hline
KDD Cup '99 \cite{protic2018review} &
1999   &
\centered{A subset of the DARPA \\ dataset used to extract \\ features} &
\centered{ Connection features (41 Features): \\  - Basic features  \\- Traffic features (same host/service) \\ - content feature } &
\centered{    - U2R \\    - R2L \\    - DoS \\    - Probe \\} 
 
\\\hline
NSL-KDD \cite{tavallaee2009detailed} &
2009   &
 \centered{Enhanced version of \\KDD Cup '99 dataset using \\ statistical analysis} &
\centered{ Connection features (41 Features): \\  - Basic features  \\- Traffic features (same host/service) \\ - content feature } &
\centered{    - U2R \\    - R2L \\    - DoS \\    - Probe \\}

\\\hline
Gas Pipeline \cite{morris2014industrial, morris2015industrial} &
2014   &
\centered {Two laboratory-scale ICS \\ - a gas pipeline \\ - a water storage tank} &
\centered{- Network traffic,\\- Process control, \\ - Process measurement features} &
\centered{ -  Reconnaissance \\   - Response Injection \\    - Command Injection \\  - Denial-of-Service}

\\\hline
UNSW-NB15 \cite{moustafa2015unsw} &
2015   &
\centered {Used IXIA PefectStorm Tool \\ to generate normal and\\ malicious  traffic} &
\centered{ Flow features (47 Features): \\  - Flow features  \\- Basic features  \\- Content features \\- Time features \\- additional general features \\ - additional connection features } &
\centered{ - Fuzzers \\   - Backdoors \\    - DoS \\  - Exploits \\    - Generic \\    - Reconnaissance \\    - Shellcode \\  - Worms \\}

\\\hline
SWaT \cite{goh2016dataset} &
2016   &
\centered { Small-scale water \\treatment system (SWaT)} &
\centered{ Packet features (18 Features):  \\- Packet basic features \\ - TCP flags \\- Modbus (code, value)  \\Process state Variables (51 features)  } &
\centered{ Multiple customized\\ attack scenarios, \\ with various attacks \\ per scenario }

\\\hline
CICIDS2017 \cite{sharafaldin2018toward} &
2017   & \centered{  Profiling the abstract \\ behaviour human users} &
\centered{ Flow features (80 Features):  \\  - Flow features  \\- Basic features  \\- Content features \\- Time features  } &
\centered{ - Brute Force FTP, SSH \\    - DoS, DDoS \\    - Heartbleed \\    - Infiltration \\    - Botnet  }

\\\hline
Electra \cite{gomez2019generation} &
2019   & \centered{ Network traffic generated \\from normal and attack \\situations at a traction\\ substation} &
\centered{ Modbus features (10 Features): \\ - Timestamp \\ - IP and MAC addresses \\ - Function code and data  \\ - Error, memory address    } &
\centered{ - False Data Injection \\    - Replay \\    - Reconnaissance }

\\\hline

TON\_IoT \cite{alsaedi2020ton_iot} &
2020   & \centered{  Telemetry data of IoT/IIoT \\ Sevices} &
\centered{ Telemetry data of Devices \\Operating Systems logs \\ Network traffic \\ - General features \\ - HTTP, DNS, SSH features  } &
\centered{ - MitM $\-\-$ - Password \\   - Scan $\-\-$ -Ransomware \\    - (D)DoS $\-\-$- Injection \\    - Backdoor $\-\-$- XSS}

\\\hline
WDT \cite{faramondi2021hardware} &
2021   & \centered{  Emulated water distribution \\testbed, Including HIL and \\controlling network} &
\centered{ Packet features (14 Features):  \\- Packet basic features \\ - TCP flags \\- Modbus (code, value)  \\Process state Variables (21 features)  } &
\centered{ - MitM \\    - Scan \\    - DoS \\    - Physical attacks}

\\\hline
ICS-Flow [This article] &
2022   & \centered{  Emulated bottle filling factory \\ testbed, Including HIL and \\controlling network} &
\centered{ Flow features (54 Features):\\  - Flow features  \\- General features  \\- TCP features \\- Extended labeling features \\ Process state variables \\ Attack logs \\ Network packets}  &
\centered{ - MitM (Injection) \\    - Scan \\    - DoS \\    - Replay Attack}
\\\hline

\end{tabular}
\end{center}
\end{table*}

Gómez et al. presented Electra\footnote{\url{http://perception.inf.um.es/ICS-datasets/}}, an anomaly detection  dataset \cite{gomez2019generation} for heterogeneous ICS scenarios. They selected the railway industry, and the Electra dataset was conducted using network traffic generated from normal and attack situations at a traction substation. Although this work has some highlighted features, such as using realistic devices (PLCs and SCADA), network protocols (Modbus), and scenarios, their extracted features are limited to Modbus-related features such as function code and errors. Therefore, this dataset lacks general network features to detect intrusion.

TON\_IoT dataset, presented by Alsaedi et al. \cite{alsaedi2020ton_iot}\footnote{\url{https://research.unsw.edu.au/projects/toniot-datasets}}, comprises Telemetry data of IoT/IIoT devices collected in a controlled environment during both normal operations and in the presence of different cyber-attacks. In addition, the dataset also includes operating systems logs (such as disk or memory usage and process information) and network traffic of an IoT network, acquired from a realistic representation of a medium-scale network at the Cyber Range and IoT Labs at the UNSW Canberra (Australia). This dataset can be employed to create and assess data-driven intrusion detection systems for IoT and IIoT environments. While this dataset includes various attack types, it falls short in terms of comprehensive network features. It solely consists of fundamental network features like the network addresses, the amount of transferred traffic, and a few protocol-specific features such as HTTP, DNS, and SSH. In order to gain a more holistic understanding of network behavior during attacks, additional enriched network features should be incorporated into the dataset. Moreover, although the TON\_IoT Dataset provides telemetry data for IoT/IIoT devices, it differs considerably from the controlling system's domain targeted by our article, where controllers (PLCs) issue controlling commands in a loop.

A Water Distribution Testbed (WDT) proposed by Faramondi et al. \cite{faramondi2021hardware} was used to generate an intrusion detection dataset for ICSs. They emulated water flowing between 8 tanks as Hardware in a Loop (HIL) and used miniCPS \cite{antonioli2015minicps} as a simulation tool to simulate the control system and networking infrastructure. This testbed had real hardwired subsystems, virtually connected to a simulated one. Multi-controller implementation is a key component of this study since it enables attacks that target communication between controllers. The proposed dataset includes both system state variables and network data to reveal attack consequences on physical processes and network traffic. This dataset has some issues, including the fact that the network dataset only contains packet-based features, which means network flow parameters are not included.

Table \ref{C:tab:comparision} summarizes the differences between the available datasets for intrusion detection experiments in industrial systems, including variations in logged data, implemented attacks, and industry domain. Furthermore, the dataset's quality can be significantly affected by the specific implementation of the testbed and attacks. Consequently, the next section offers more information on the development of the ICS-Flow dataset to illustrate its potential value as a resource for evaluating industrial IDSs.

\section{ICS-Flow Dataset}
\label{C:sec:mainwork}

In this section, we introduce the ICS-Flow dataset, which is designed as an evaluation dataset for ML-based IDSs intended for industrial systems. We first present our industrial testbed environment. Then, we describe selected attack types and explain how the attacks are implemented. Next, we detail the creation of the intrusion detection dataset, including the network features derived and the network packet processing techniques used to generate network flow features. Finally, we introduce the labeling process and our different strategies for labeling.

\subsection{Testbed Environment}

\begin{figure}[hb!]
\centering
        \includegraphics[width=0.8\linewidth]{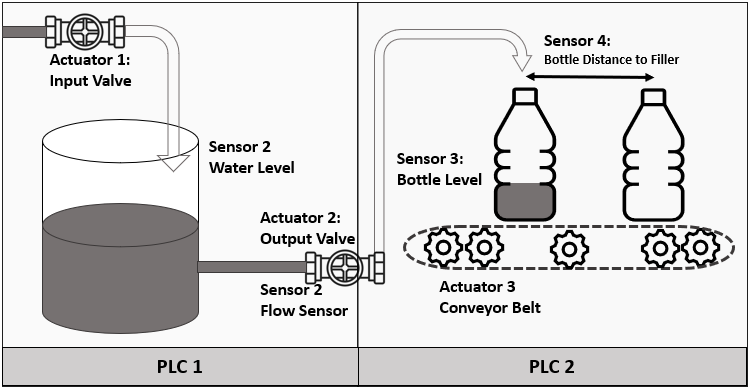}
        \caption{The Sample bottle filling factory presented in ICSSIM framework \cite{dehlaghi2023icssim}}
        \label{C:fig:physical_process}
\end{figure}

We used the bottle filling factory simulation provided by ICSSIM \cite{dehlaghi2023icssim} as a virtual testbed to investigate cyber threats and attacks. The simulation, which is illustrated in Figure \ref{C:fig:physical_process}, includes an ICS that controls the equipment within the factory - such as pipes, valves, a conveyor belt, and a water tank - to fill empty bottles with water from the tank. The input valve regulates the water level in the tank to ensure that it remains within the permissible range, while the output valve controls the flow of water for filling bottles. The conveyor belt engine ensures empty bottles are positioned correctly beneath the filler. PLC-1 reads the tank water level and pipe water flow sensors and sends control commands to turn input and output valves \textit{`On'} and \textit{`Off'}. PLC-2 monitors the water level in filling bottles and the distance between the filler and the next bottle using sensors. It also issues control commands for the conveyor belt based on sensor readings and communications with PLC-1.

\begin{figure}[hb!]
\centering
        \includegraphics[width=0.8\linewidth]{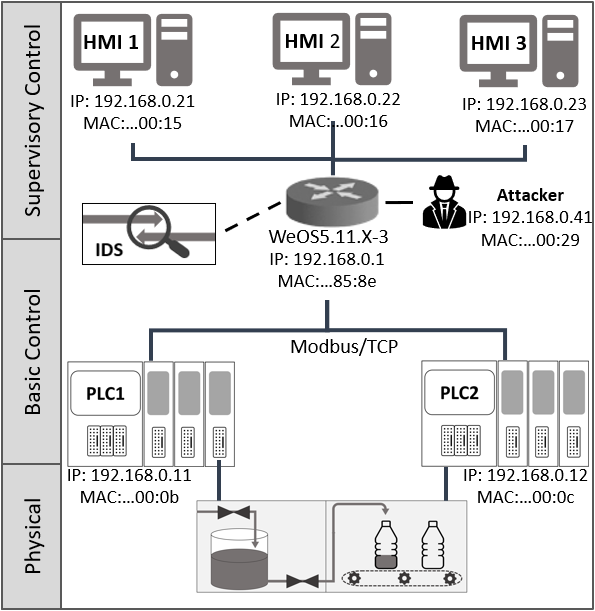}
        \caption{The testbed architecture}
        \label{C:fig:architecture}
\end{figure}

We enhanced the virtual testbed from \cite{dehlaghi2023icssim} by introducing new types of  Human Machine Interfaces (HMI), control logic, and attacks. The final architecture of the testbed, depicted in Figure \ref{C:fig:architecture}, consists of two PLCs that manage a bottle filling factory and three HMIs that supervise the system and send manual operational commands. All the components of the ICS are connected through a network, and Table \ref{C:tab:configurations} provides detailed network configurations of the ICS nodes. In this testbed, all ICS components operate on Docker containers\footnote{\url{https://www.docker.com/}} in automatic mode since the simulation is conducted using ICSSIM. However, the use of automated HMIs in this testbed may create uniform network traffic during regular operation, which could decrease the difficulty of anomaly detection in the dataset. To address this concern, we added random parameters to the HMIs' behavior.
\begin{itemize}
    \item HMI1: constantly reads all controlling signals. 
    \item HMI2: sends write commands based on a predefined scenario to simulate HMIs' user commands. 
    \item HMI3: sends eligible write commands with uniform random values.
\end{itemize}

\begin{table}[]
\caption{ICS nodes network configurations in our testbed}
\label{C:tab:configurations}
\begin{center}
\begin{tabular}{ |p{1.2cm}|p{2.5cm} |p{2cm}| }
 \hline
 \textbf{Name}   &   \textbf{Mac}   &   \textbf{IP}     \\\hline
 
PLC1 &
02:42:c0:a8:00:0b & 
192.168.0.11 
\\\hline
PLC1 &
02:42:c0:a8:00:0c & 
192.168.0.12 
\\\hline
HMI1 &
02:42:c0:a8:00:15 & 
192.168.0.21 
\\\hline
HMI2 &
02:42:c0:a8:00:16 & 
192.168.0.22 
\\\hline
HMI3 &
02:42:c0:a8:00:17 & 
192.168.0.23 
\\\hline

Attacker &
02:42:c0:a8:00:29 & 
192.168.0.41 
\\\hline

\end{tabular}
\end{center}
\end{table}

In order to simulate network communication realistically, the Modbus protocol has been employed for all communications. Modbus is a communication protocol widely used in operational ICSs \cite{parian2020fooling}, originally designed for PLCs and later adopted as the de facto standard for connecting industrial electronic devices \cite{thomas2008introduction}. We use Modbus TCP to ensure compatibility with a variety of industrial devices. PLCs function as Modbus-TCP servers, while ICS components establish communication by creating Modbus-TCP client instances that communicate with Modbus-Servers. 

To launch cyberattacks against the testbed, we connect an attacker node to the ICS network. It is assumed that the attacker has access to internal network traffic because many ICS systems now offer remote monitoring for industrial equipment, allowing attackers to access the ICS internal network through infected remote HMIs. However, the attacker does not have prior knowledge of the control system or controlling variables. Therefore, the attacker must sniff the network and gather the necessary information for complex attacks.


After setting up the testbeds, we captured network data to create the ICS-Flow dataset, which includes the communication between ICS components. We captured network packets by running TCPdump on the switch, and this data can be employed to train IDSs on the switch or as a standalone component within the control network. Implementing the intrusion detection module within the control network facilitates the detection of any dubious control activities and allows for the monitoring of traffic among ICS components. This is particularly useful in identifying attacks that launched internally or attacks that have already penetrated the network perimeter.

\subsection{Attacks}
\label{C:subsec:attacks}

Protocols used in ICSs are vulnerable due to their lack of authentication, communication encryption, and integrity checks \cite{gomez2019generation}. These weaknesses allow attackers to eavesdrop on or alter network transmissions, potentially disrupting ICS operations. To exploit these vulnerabilities, we considered four attack types to discover the network, sniff or modify packets and disrupt ICS operation. We selected the attacks regarded as common attack types in previous studies \cite{faramondi2021hardware,morris2014industrial,gomez2019generation}. Moreover, the  \textit{'MITRE ATT\&CK'}\footnote{\url{https://attack.mitre.org/techniques/ics/}}, which is a globally-accessible knowledge base of adversary techniques based on real-world observations, and European Union Agency for Cybersecurity (ENISA) reports\footnote{\url{https://www.enisa.europa.eu/publications/good-practices-for-security-of-iot}} \cite{enisa2018good} consider these attacks to be common cyberattacks on ICSs.

Our team has developed a  Python script that automates the execution of attacks on the control system. Using this script, the attacker node can quickly launch attacks based on a predefined scenario. We have also designed the script to allow the control system sufficient time to recover from any destabilization caused by the attacks, ensuring that the system remains stable and secure. The details of implemented attacks are listed below.

\subsubsection{Reconnaissance Attack}

Intruders often begin with a reconnaissance attack, which involves gathering valuable system information as a preliminary step towards launching further attacks \cite{wang2022cyber}. A reconnaissance attack is passive since the attacker merely captures information rather than disrupting the victim's functionalities. This attack does not impact industrial operations, although its trace on the network traffic can still reveal it. In our implementation, the intruder discovers the network by collecting information such as IP addresses, media access control addresses (MAC), or open ports.

We have implemented the Reconnaissance attack with two different scenarios: IP-Scanning and Port-Scanning. In the first scenario, we utilize Ettercap\footnote{\url{https://www.ettercap-project.org/}}, a free and open-source network security tool, and a Python script using the Scapy\footnote{\url{https://scapy.net/}}, a packet manipulation tool for networks, to broadcast Address Resolution Messages (ARP) to discover the alive network nodes.  In the second scenario,  we use NMap\footnote{\url{https://nmap.org/}}, a free and open-source utility for network discovery, to gather information about the hosts and ports on them in the network.  NMap uses a port scanning technique to find vulnerabilities on victim hosts.  Each scenario could reveal information about the target network with different traces on the dataset. 
    
\subsubsection{Replay Attack}
A commonly used attack in control systems is the replay attack, which involves exploiting valid network packets captured during normal system operation by maliciously retransmitting them \cite{gomez2019generation}. In this type of attack, the attacker sniffs the network passively to collect valid packets and then actively sends the recorded packets frequently to other nodes, disrupting the system's normal behavior. Since replaying valid packets at inappropriate times can lead to unexpected results, this attack does not require in-depth knowledge of network traffic packets. It is essential to note that this type of attack can be highly damaging, even with minimal information about the system.

We have implemented the replay attack using Scapy by developing a Python script that executes the attack in two phases. In the first phase, the attacker uses ARP poisoning and the MitM technique to sniff network packets for 15 seconds. These packets are later replayed three times (45 Seconds) by the attacker to disrupt the control system's regular operation.  However, replaying the exact same messages is not possible since every TCP connection depends on two  32-bit random sequence numbers generated by the client/server, and packets with duplicate sequence numbers will be rejected. As a solution, a replay attack on a TCP connection reuses only TCP payloads. Therefore, the attacker uses intercepted IP addresses, ports, Modbus commands, and arguments to create a new TCP-Connection to perform a replay attack.

\subsubsection{Distributed Denial of Service (DDoS) Attack}
In order to disrupt the industrial operation, attackers can flood the network or service with a large number of packets or service requests, resulting in a denial of service on the victim component \cite{ylmaz2018cyber}. Meanwhile, using multiple attackers makes the attack more effective. The DDoS target can be limited bandwidth, storage, or computing power.

We leverage the network addresses collected during the reconnaissance attack and the Modbus addresses gathered by sniffing the network to conduct a DDoS attack. In this attack, we use the 'DDoSAgent' class from ICSSIM to generate a massive flood of reading requests to the PLCs. Our implementation involves a script that spawns 800 instances of the 'DDoSAgent' class, which relentlessly bombards the PLCs with read requests for 60 seconds. This results in significant delays in the ICS network communication during the attack.
    
\subsubsection{MitM Attack}
We use a MitM attack to inject false data into the controlling system. The false data injection attack is one of the most critical
malicious cyberattack \cite{reda2022comprehensive} that involves injecting incorrect data into the controlling system, compromising the communication in ICSs.
Using the MitM technique, an attacker intercepts or manipulates communications between two ICS components while they believe they are interacting directly with each other \cite{lan2020traffic}. As part of our testbed, we employ MitM techniques to inject false data into the ICS system. Therefore, we refer to this attack as the false data injection attack and the MitM attack.

To execute this attack, the attacker first performs ARP poisoning on the ICS component, redirecting network packets to their node. Next, the attacker intercepts the packets and modifies the Modbus write requests and read responses by multiplying their values with a specific factor. The attacker then sends the manipulated packets to the intended destination. Finally, ARP messages are sent to the ICS component to clear routing tables and erase any evidence of the attack. A custom Python script based on the Scapy library was used to carry out the attack. The attack was repeated several times for a duration of 30 seconds each time, resulting in a 10\% error in the Modbus read response data.

\subsection{Feature extraction}
Existing literature in the network security domain captures the network either in packet-based or flow-based format \cite{ring2019survey}. Although packet-based datasets provide detailed information regarding network anomalies, AI analysis of these voluminous datasets is time-consuming, making it challenging to develop an IDS. An alternative is using the network flows format, which aggregates packets in a time interval with certain properties. The typical shared properties are network protocol, Source IP, Destination IP, Source Port, and Destination Port \cite{claise2008specification}. The flow definition varies depending on the testbed configuration. We also use a customized definition of flow since our testbed uses the Modbus protocol, which always uses port number 502 as the default port. Therefore, we define a flow as an aggregated record of network packets within a time interval with a similar source address, a destination address, and network protocol. Based on the time setting of our testbed, we considered 500 ms as an interval for flows in this experiment.

To generate a network flow dataset, we developed the ICSFlowGenerator tool using the Scapy library, which is available in our public GitHub repository\footnote{\url{https://github.com/AlirezaDehlaghi/ICSFlow}}. The ICSFlowGenerator tool follows the procedure presented in Algorithm \ref{C:alg:flowgenerator}. It iterates through network packets in a dumped PCAP file and generates flow features based on the interval option. The output is a flow file in CSV format, containing flow-based features. For network flows that are not TCP-based, the TCP payload features columns are empty. In addition to TCP traffic, our dataset includes ARP messages to detect MitM attacks. To account for the lack of IP addresses in ARP messages, we generate ARP flows based on their MAC addresses.

\begin{algorithm}

\caption{- ICSFlowGenerator}\label{C:alg:flowgenerator}
\begin{algorithmic}[1]
\Statex \textbf{Inputs:} $\textit{PcapFile, Interval}$
\Statex \textbf{Output:} $\textit{FlowFile}$
\Procedure{GenerateFlows}{}
\State $flows \gets \varnothing$ 

\State \textbf{foreach} $\textit{packet}, \textit{time}$ \textbf{in} $\textit{PcapFile}$
\Statex  \hspace{1.2cm} {\textcolor{xanadu}{// discard LCC frames without protocol}}
\State \hspace{0.7cm} \textbf{if} {$`type`$ \textbf{not in} $\textit{packet.fields}$ \textbf{then}} 
\State \hspace{1.4cm} \textbf{Conitnue} 
\State \hspace{0.7cm} $src \gets \textit{Min}(packet.src, packet.dst)$
\State \hspace{0.7cm} $dst \gets Max(packet.src, packet.dst)$
\State \hspace{0.7cm} $protocol \gets packet.protocol$
\State \hspace{0.7cm} \textbf{if} $flows[src, dst, protocol]$ \textbf{then}
\State \hspace{1.4cm} $flow \gets flows[src, dst, protocol]$ 
\State \hspace{0.7cm} \textbf{else} 
\Statex  \hspace{1.9cm} {\textcolor{xanadu}{// a flow is a collection of packets}}
\State \hspace{1.4cm} $flow \gets new$ $flow(src, dst, protocol)$ 
\Statex  \hspace{1.2cm} {\textcolor{xanadu}{// flush flows that was open for windows lenght}}
\State \hspace{0.7cm} \textbf{if} $time-flow[0].time>Interval$ \textbf{then}
\State \hspace{1.4cm} \textbf{Print($flow$) in $FlowFile$ }
\State \hspace{1.4cm} $flow \gets \varnothing$
\State \hspace{0.7cm} $flow.add(packet, time)$ 
\State \hspace{0.7cm} $flow.UpdateFeatures()$ 
\State \textbf{foreach} $\textit{flow}$ \textbf{in} $\textit{flows}$
\Statex  \hspace{1.2cm} {\textcolor{xanadu}{// flush rest of flows}}
\State \hspace{0.7cm} \textbf{Print($flow$) in $FlowFile$ }

\EndProcedure
\end{algorithmic}
\end{algorithm}

The flow dataset contains 54 columns, including 50 features and 4 label columns. Feature columns are categorized into three categories: flow features, general features, and TCP features. Table \ref{C:tab:flow_features} presents the flow features, including the source address, destination address, and network protocol. Table \ref{C:tab:general_features} introduces the general features category with 16 features that are shared for all network flows regardless of the protocol type. This dataset also contains 20 TCP header features outlined in Table \ref{C:tab:tcp_features}. These features are extracted from TCP headers and contain statistics about Flags, time to live, TCP window, and delays.

\begin{table}[]
\centering
\caption{Flow features} 
\label{C:tab:flow_features}
\begin{tabular}{ |p{0.7cm}|p{2cm}|p{3.9cm} |p{0.5cm}|}
 \hline
 \textbf{\#}   &\textbf{Feature}   &   \textbf{Description}   &   \textbf{Type}    \\\hline
1, 2& (s/r)Address & Sender/Receiver IP address or MAC address of flow & str \\\hline
3& protocol & Packet type and Network Protocol (ARP, IPV4-TCP, IPV4-UDP, ...) & str \\\hline
\end{tabular}
\end{table}

\begin{table}[]
\centering
\caption{General features }
\label{C:tab:general_features}
\begin{tabular}{ |p{0.7cm}|p{2cm}|p{3.9cm} |p{0.5cm}|}
 \hline
  \textbf{\#}   &\textbf{Feature}   &   \textbf{Description}   &   \textbf{Type}    \\\hline
4& start & Time stamp of the first packet in the flow & float \\\hline
5& end & Time stamp of the last packet in the flow & float \\\hline
6& startOffset & Time offset of first packet in the flow relative to first packet in the dataset & float \\\hline
7& endOffset & Time offset of last packet in the flow relative to first packet in the dataset& float \\\hline
8& duration & Flow interval, the difference between end and start time stamps & float \\\hline
9, 10& (s/r)Packets & Sent or received packets count within the flow & int\\\hline
11, 12& (s/r)BytesMax & Maximum bytes sent or received in one packet within the flow & int\\\hline
13, 14& (s/r)BytesMin & Minimum bytes sent or received in one packet within the flow & int\\\hline
15, 16& (s/r)BytesAvg & Average sent or received bytes per packet within the flow & float\\\hline
17, 18& (s/r)Load &  Sent or received bits per second & float  \\\hline
19, 20& (s/r)PayloadMax & Maximum bytes sent or received as a payload in one packet within the flow & int\\\hline
21, 22& (s/r)PayloadMin & Minimum bytes sent or received as a payload in one packet within the flow & int\\\hline
23, 24& (s/r)PayloadAvg & Average payload bytes of sent or received packets & float\\\hline
25, 26& (s/r)InterPacket &  Average send or receive inter-packet arrival time & float  \\\hline

\end{tabular}
\end{table}

\begin{table}[]
\centering
\caption{TCP features }
\label{C:tab:tcp_features}
\begin{tabular}{ |p{0.7cm}|p{2cm}|p{3.9cm} |p{0.5cm}|}
 \hline
  \textbf{\#}   &\textbf{Feature}   &   \textbf{Description}   &   \textbf{Type}    \\\hline
27, 28& (s/r)ttl & Average sending or receiving time-to-live & float \\\hline
29, 30& (s/r)AckDelayMax & Maximum interval between packets and their acknowledge in sender or receiver node & float \\\hline
31, 32& (s/r)AckDelayMin & Minimum interval between packets and their acknowledge in sender or receiver node & float \\\hline
33, 34& (s/r)AckDelayAvg & Average interval between packets and their acknowledge in sender or receiver node & float \\\hline

35, 36& (s/r)AckRate & Rate of sent or received packets contains \'acknowledge\' flag & float \\\hline
37, 38& (s/r)FinRate & Rate of sent or received packets contains 'Finish' flag & float \\\hline
39, 40& (s/r)PshRate & Rate of sent or received packets contains 'Push' flag & float \\\hline
41, 42& (s/r)RstRate & Rate of sent or received packets contains 'Reset' flag & float \\\hline
43, 44& (s/r)UrgRate & Rate of sent or received packets contains 'Urgent' flag & float \\\hline
45, 46& (s/r)AckRate & Rate of sent or received packets contains 'Synchronisation' flag & float \\\hline
47, 48& (s/r)WinTCP &sender or receiver average TCP window advertisement & float \\\hline
49, 50& (s/r)FragmentRate &sender or receiver Fragmentation rate of TCP packets & float \\\hline

\end{tabular}
\end{table}

\subsection{Labelling}
The goal of labeling is to provide information about the context of data. Although the quality of labelling data is not precisely defined, several articles have stressed the importance of accurate labeling as a crucial component of producing high-quality network traffic datasets \cite{macia2018ugr}. Human-guided or automatic labeling are two possible strategies to label datasets. In this study, automatic labeling was employed since it is fast, requires little expertise, and is easy to adapt for all types of attacks \cite{guerra2022datasets}. The data required for training ML models are different depending on how we define ML-based intrusion detection. In this article, we aim to support two types of definitions, anomalous record detection, and intrusion detection. 

In anomalous record detection, the network trace dataset is labeled as normal or malicious, and the ultimate goal is to classify dataset records into different categories. Several approaches exist to do this classification, including binary classifications for attack detection or multi-class classifications for attack identification. However, the labeled dataset is always required as a training set in supervised approaches or an evaluation set for unsupervised ML methods, as record labels are available in some datasets such as \cite{tavallaee2009detailed, moustafa2015unsw,sharafaldin2018toward, gomez2019generation}. Automatically recognizing anomalous network traces from normal traffic is challenging \cite{guerra2022datasets}. To overcome this, we use two different strategies for labeling; namely, Injection Timing (IT) \cite{lemay2016providing} and Network Security Tools (NST) \cite{guerra2022datasets}. In the former strategy, we consider all network traffic during an attack anomalous, while in the latter strategy, we only consider network traffic coming from or going toward the attacker node as an anomaly.

In intrusion detection, the ultimate objective is to find attack occurrences by analyzing unlabeled datasets and system logs. In other words, we do not look for anomalous flows but rather for anomalous system behavior in a time period. A log data of attack occurrence, along with unlabeled network and system data, is required for this task to verify the intrusion detection results. As pointed out in \cite{goh2016dataset}, a combination of unlabeled network traces and the attack log file forms an unsupervised or semi-supervised dataset for attack detection.

By providing attack occurrence log files and labeled network traces using IT and NST strategies, we provide a dataset with the flexibility to perform all described anomaly detections. The labelling features are listed in table \ref{C:tab:label_features}. However, we believe that unlabeled datasets are more realistic for industrial implementation since the final mission of IDSs is attack detection, and labeled flows are not available in operational ICSs. 

Finally, as a complement to the class labels using the IT and NST labeling strategy in binary and multi-class values, we present the attack log file in Table \ref{C:tab:logfile} to provide a ground for further analysis.

\begin{table}[]
\centering
\caption{Label features }
\label{C:tab:label_features}
\begin{tabular}{ |p{0.7cm}|p{2cm}|p{3.9cm} |p{0.5cm}|}
 \hline
  \textbf{\#}   &\textbf{Feature}   &   \textbf{Description}   &   \textbf{Type}    \\\hline
51& IT-B-Label & 0 if record is normal, 1 if record is malicious (using IT labeling methodology) & int \\\hline
52&  IT-M-Label & 'normal' if record is normal, 'attack-name' if record is malicious (using IT labeling methodology) & str\\\hline
53& NST-B-Label & 0 if record is normal, 1 if record is malious (using NST labeling methodology) & int \\\hline
54& NST-M-Label & 'normal' if record is normal, 'attack-name' if record is malicious (using NST labeling methodology) & str\\\hline
\end{tabular}
\end{table}

\begin{table}[]
\begin{center}
\caption{Attack log file }
\label{C:tab:logfile}
\begin{tabular}{ |p{2.0cm}|p{5.8cm} |  }
 \hline
 \textbf{Field Name}   &   \textbf{Description}     \\\hline
 
\textbf{Attack} &
Shows the attack that applied in the dataset\\\hline

\textbf{Start time} &
Attack start time stamp in Unix time format\\\hline

\textbf{End time} &
Attack end time stamp in Unix time format\\\hline

\textbf{Attacker IP} &
IP address of the attacker\\\hline

\textbf{Attacker MAC} &
IP address of the attacker\\\hline

\textbf{Extra Info} &
Contains extra information about the applied attack 
\\\hline

\end{tabular}
\end{center}
\end{table}

\section{Dataset Analysis }
\label{C:sec:datasetmetrics}
This section will provide a detailed account of how we implemented our experiment and conduct a thorough analysis of the ICS-Flow dataset. The analysis will include statistical information about the dataset records, an explanation of the final format of the dataset files, and using 2D representations to visualize the dataset.

The network configuration was optimized to accommodate the burst of network packets caused by a DDoS attack. To capture all packets, we increased the buffer size of the Switch to 400MB and utilized TCPdump to capture traffic. The testbed was operational for three hours, with the first hour being attack-free. This period without attacks provided normal samples for anomaly detection techniques that use semi-supervised AI. Over the following two hours, attacks were conducted randomly and intermittently. We included gaps between attacks to prevent attack overlaps, considering the time required for system recovery after each attack.

\begin{figure}[hb!]
\centering
        \includegraphics[width=0.9\linewidth]{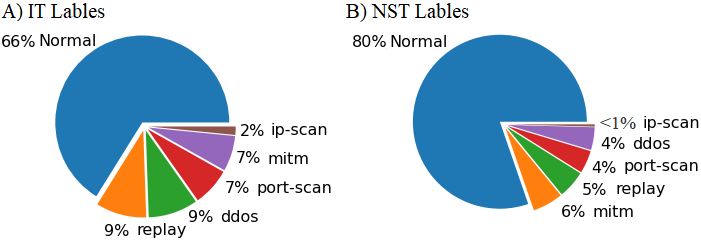}
        \caption{IT and NST Class labels distribution over ICS-Flow dataset.}
        \label{C:fig:labels}
\end{figure}

We captured a total of 2GB of raw network traffic, comprising over 25 million packets. Subsequently, we analyzed this traffic using ICSFlowGenerator and generated a final dataset of 45719 network flows. The detailed statistics on the labeled flows using IT and NST labeling strategies are presented in Table \ref{C:tab:lables} and Figure \ref{C:fig:labels}. These statistics reveal that the number of attack flows identified using the NST approach is lower than that of the IT approach, as the NST method has stricter criteria for classifying a flow as an attack. Furthermore, as part of the experiment, we logged process variables using the PLCs' logger. These log files are available in CSV format and can aid researchers in identifying anomalies based on process variable analysis. Lastly, the dataset comprises four files, namely, the raw Pcap file, the labeled network flow dataset, the attack log file, and process state variables, all of which are publicly accessible in our repository\footnote{\url{https://www.kaggle.com/datasets/alirezadehlaghi/icssim}}.
 
\begin{table}[]
\centering
\caption{Flow statistics using IT and NST labeling strategies in ICS-Flow dataset }
\label{C:tab:lables}
\begin{tabular}{|l|l|l|}
\hline
\textbf{Attack type} & \textbf{\# of IT Flows} & \textbf{\# of NST Flows} \\ \hline
Normal               & 30236                   & 36706                    \\ \hline
IP-Scan              & 712                     & 192                      \\ \hline
Port-Scan            & 3235                    & 1944                     \\ \hline
Replay               & 4300                    & 2358                     \\ \hline
DDoS                 & 4221                    & 1934                     \\ \hline
MitM                 & 3014                    & 2584                     \\ \hline
Total                & 45719                   & 45719                    \\ \hline
\end{tabular}
\end{table}

\begin{figure*}[t!]
\centering
        \includegraphics[width=0.8\linewidth]{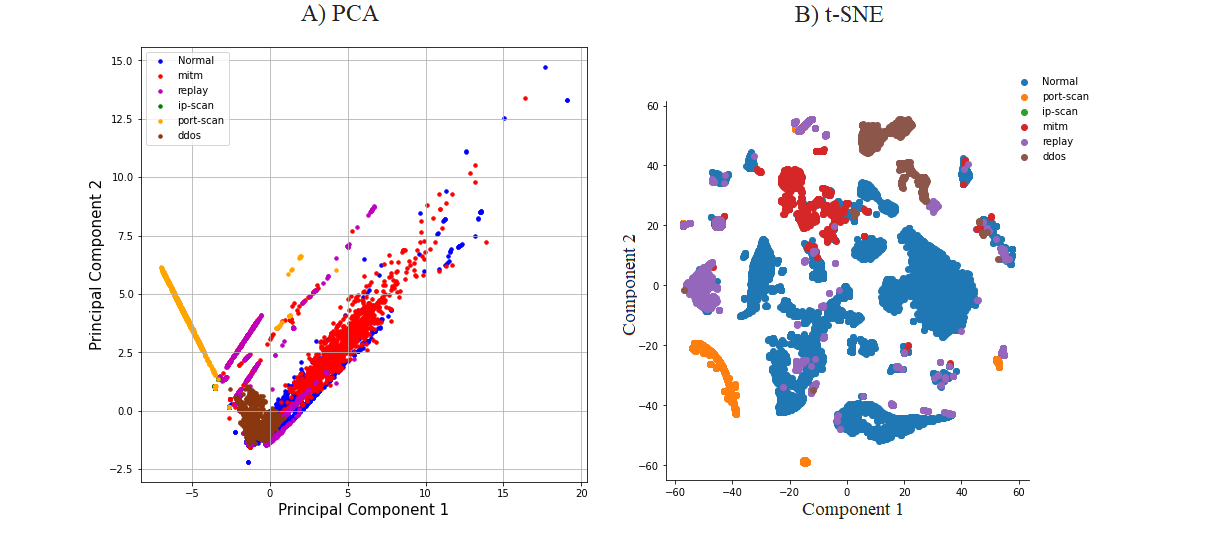}
        \caption{2D presentation of ICS-Flow dataset records using PCA and t-SNE techniques}
        \label{C:fig:presentation}
\end{figure*}

Visualizing dataset records help us identify data patterns. However, a large number of normal network flows in this unbalanced dataset obscure the attack network flows. To alleviate this problem, we chose to visualize only the second half of the dataset, which contains a mix of normal and attack records. We applied two techniques to show multi-dimensional data in the 2D diagrams. Firstly, we applied Principal Component Analysis (PCA) \cite{wold1987principal}, a popular technique for analyzing large datasets with many features. Although PCA is primarily used for reducing data dimensions, extracting the first two or three features enables us to demonstrate data in 2D or 3D presentation. Figure \ref{C:fig:presentation}-A demonstrates that while PCA is helpful for visualizing data, it does not clearly show data clusters. Secondly, to visualize the high-dimensional data, we employed t-Distributed Stochastic Neighbor Embedding (t-SNE) \cite{van2008visualizing}, a statistical method known for its effectiveness in dimensionality reduction. We conducted a grid search over a range of perplexity values [30, 50, 100, 250, 500, 1000], and chose a perplexity of 250 based on its optimal performance. Figure \ref{C:fig:presentation}-B illustrates the 2D presentation of the ICS-Flow dataset using t-SNE with perplexity 250. Our analysis revealed that DDoS, Port-Scanning, and MitM attacks form distinct clusters. However, distinguishing between replay attacks and normal attacks is challenging since replay attacks are designed to mimic the behavior of a normal system. Furthermore, IP-Scan attack records are merged with other groups,  which we will analyze and explain in the results and discussion section.  

\section{ML-Based Anomaly Detection Models}
\label{C:sec:ai}
\begin{figure}[hb!]
\centering
        \includegraphics[width=\linewidth]{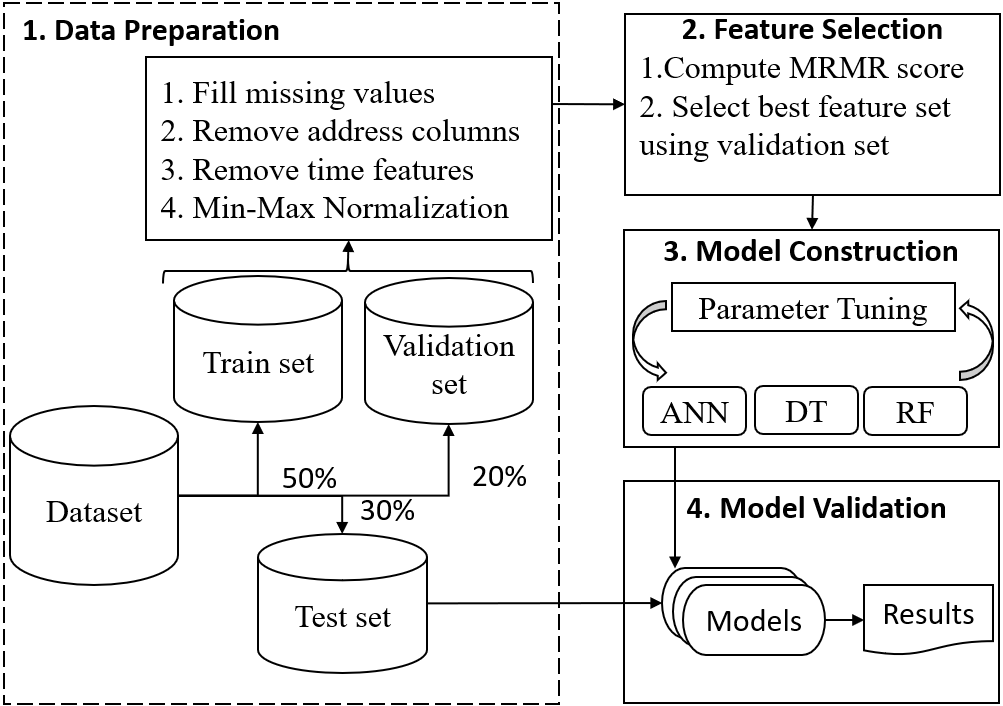}
        \caption{Methodology to assess ML-methods}
        \label{C:fig:methodology}
\end{figure}

This section illustrates how ML techniques can be applied to detect and identify intrusions using our dataset. We conducted two separate experiments with the dataset: intrusion detection and identification. The objective of the intrusion detection experiment was to identify attack flows irrespective of their attack type. To achieve this, we classified records into normal or attack flows. In the intrusion identification experiment, we aimed to identify the specific attack type, which could help devise an effective mitigation strategy. The performance evaluation of the ML methods on the 'ICS-Flow' dataset was conducted using a four-phased methodology, depicted in Figure \ref{C:fig:methodology}. In the rest of this section, we will talk about actions performed in these phases in detail.

\subsection{Data Preparation}
During the dataset preprocessing phase, the following steps
were conducted:
\begin{enumerate}
    \item  In the dataset, certain rows have missing column values due to the inability to compute TCP features for non-TCP protocol flows like UDP flows. To address this, we have replaced the missing values with a placeholder value of '0'.
    \item For the attack detection and identification experiments, the `NST-B-Label' and `NST-B-Label' columns were chosen as output variables, respectively.
    \item The sender address (`sAddress') and receiver address ('rAddress') columns were excluded from the experiment to prevent the attack detection task from becoming trivial due to the use of a fixed IP for the attacker.  
    \item The start time (`start' and `startOffset') and end time (`end' and `endOffset') columns were eliminated, as they solely entail temporal data and do not offer any valuable understanding of the cyberattacks.
    \item The dataset was partitioned into three subsets: training, validation, and test sets, which comprised 50\%, 20\%, and 30\% of the data, respectively.
    \item We utilized the `Min-Max' normalization method described in Eq. \ref{C:eq:normalization} to normalize the numerical features. Here, Min and Max represent the minimum and maximum values of a feature in the training set, respectively.
\end{enumerate}

\begin{equation}
\label{C:eq:normalization}
  V_{Norm} = \frac{V - Min}{Max - Min}
\end{equation}

\begin{table}[]
\centering
\caption{MRMR score of selected feature}
\label{C:tab:selected_features}
\begin{tabular}{ |p{1.2cm}|p{4.0cm}|p{2.1cm} |}
 \hline
 \textbf{Rank}   &\textbf{Feature}   &   \textbf{MRMR Score} \\\hline
1 \& 2   &   rBytesAvg \& sBytesAvg        & 0.4309 - 0.0559 \\\hline
3 \& 4   &   sFinRate \& rFinRate      &0.4058 - 0.1479 \\\hline 
5 \& 6   &   sSynRate \& rSynRate     &0.3886 - 0.1731 \\\hline 
7 \& 8   &   sRstRate \& rRstRate   & 0.2323 - 0.2072 \\\hline 
9 \& 10  &   rttl  \& sttl        &0.1738 - 0.1075 \\\hline 
11 \& 12 &   sAckRate \& rAckRate   & 0.1366 - 0.0690 \\\hline 
13 \& 14 &   sMaxAckDelay \& rMaxAckDelay & 0.1241 - 0.0831 \\\hline 
15 \& 16 &   sPackets \& rPackets & 0.0825 - 0.0587 \\\hline 
17       & Protocol                 & 0.821 \\\hline 
18 \& 19 & sWinTCP \& rWinTCP       & 0.0718 - 0.0688 \\\hline 
20 \& 21 & rPayloadAvg \& sPayloadAvg    & 0.0716 - 0.0671 \\\hline 
22 \& 23 & rInterpacket \& sInterpacket & 0.0701 - 0.0553 \\\hline 

\end{tabular}
\end{table}

\subsection{Feature Selection}
During the feature selection phase, we utilize the Maximum Relevance, and Minimum Redundancy (MRMR) technique \cite{peng2005feature} to reduce the dimensionality of our data. MRMR assigns a score to each feature based on its relevance and redundancy with other feature columns. To constrain the size of our feature set, we only select features with scores above 0.07. This threshold was determined by iteratively testing various thresholds and measuring the model's accuracy on the validation set. Note that many of our parameters, such as `sBytesAvg' and `rByteAvg', are dual counterparts; Therefore, selecting one requires selecting its corresponding dual parameter. Table \ref{C:tab:selected_features} presents the 23 features that have been selected for our experiment, along with their corresponding MRMR scores.  

\subsection{Model Construction}
In the third phase, we apply supervised ML models to perform the flow classification. During this phase, we use three off-the-shelf classification techniques, which have shown better performance in previous studies \cite{singh2022survey,varghese2022digital}: Decision Tree (DT), Random Forest (RF), and Artificial Neural Networks (ANN). We also optimize the accuracy of these algorithms by giving effective values to their hyper-parameter settings and measuring their performance on the validation set. Table \ref{C:tab:search-space-hyper-parameters} summarizes the examined hyper-parameters for each approach.

\begin{table}[]
\centering
\caption{Hyper-parameters search space of ML algorithms }
\label{C:tab:search-space-hyper-parameters}
\begin{tabular}{|p{0.9cm}|p{7.0cm}|}
\hline
\textbf{Method} & \textbf{Search space parameters}                                                                          \\ \hline
DT     & Split criterion: {[}Gini's diversity index, Twoing rule, and Maximum deviance reduction{]}       \\ \hline
RF     & Number of learners: {[}10:500{]}, Number of predictors to sample: {[}1:40{]}                     \\ \hline
ANN    & Layers: {[}1, 2, 3{]}  Layer size: {[}1 : 300{]} Activation: {[}ReLU, Tanh, None, and Sigmoid{]} \\ \hline
\end{tabular}
\end{table}


\subsection{Model Validation}
We chose not to balance the classes in the dataset that we generated, as we were unable to identify a realistic distribution of attacks in operating ICSs. Nevertheless, to handle the unbalanced data, we employed performance metrics that could effectively measure the model's precision on such data. Hence, in addition to accuracy, we employed metrics such as  \emph{accuracy}(Eq. \ref{C:eq:accuracy}) , \emph{recall} (Eq. \ref{C:eq:recall}), \emph{precision} (Eq. \ref{C:eq:precision}),  and \emph{F1-score} (Eq. \ref{C:eq:fscore}), and confusion matrix to showcase how well the ML methods performed on unbalanced data.

\begin{equation}
\label{C:eq:accuracy}
  Accuracy = \frac{TP+TN}{TP+FP+TN+FN}
\end{equation}

\begin{equation}
\label{C:eq:recall}
  Recall = \frac{TP}{TP+FN}
\end{equation}

\begin{equation}
\label{C:eq:precision}
  Precision = \frac{TP}{TP+FP}
\end{equation}

\begin{equation}
\label{C:eq:fscore}
  F1-Score = \frac{2}{\frac{1}{Precision} + \frac{1}{Recall} }
\end{equation}

Where $FP$=False Positive, $FN$=False Negative, $TP$=True Positive, and $TN$=True Negative.

\section{Results and Discussion}
\label{C:sec:results}
The results for ML-based intrusion detection using the optimal hyper-parameter setup (Table \ref{C:tab:hyper-parameters}) are shown in Table \ref{C:tab:results-binaryclass}. While all algorithms identify attack flows with greater than 99.4\% accuracy, the RF technique outperforms the others with 99.5\% accuracy and a higher F1-score. Despite the 98.2\% precision of the RF algorithm in detecting attacks, even a low number of false alarms can impose high disruption costs in system operation.

\begin{table}[]
\centering
\caption{Hyper-parameters of ML algorithms }
\label{C:tab:hyper-parameters}
\begin{tabular}{|p{0.7cm}|p{1.3cm}|p{4.9cm}|}
\hline
Model                & Mode        & Hyper Parameters                                             \\ \hline
\multirow{2}{*}{DT}  & Binary      &  Split criterion: Maximum deviance reduction, Maximum number of Splits: 314                                                             \\ \cline{2-3} 
                     & Multi-class & Split criterion: Twoing rule, Maximum number of Splits: 1000 \\ \hline
\multirow{2}{*}{RF} & Binary      &  Number of learners: 10, Maximum number of splits	850, Number of predictors to sample: 17                                                            \\ \cline{2-3} 
                     & Multi-class  & Number of learners: 54, Maximum number of splits:1680, Number of predictors to sample: 8                                                             \\ \hline
\multirow{2}{*}{ANN} & Binary      &  Fully connected layers: 1, Activation: Sigmoid, First layer size: 79                                                           \\ \cline{2-3} 
                     & Multi-class  & Fully connected layers:1, Activation: Sigmoid, First layer size	257                                                            \\ \hline
\end{tabular}
\end{table}

\begin{table}[]
\centering
\caption{Test results of three ML-methods for attack detection }
\label{C:tab:results-binaryclass}
\begin{tabular}{|ll|l|l|l|l|}
\hline
\multicolumn{2}{|l|}{}                                  & \textbf{Accuracy}                  & \textbf{Precision} & \textbf{Recall}    & \textbf{F1-score} \\ \hline
\multicolumn{1}{|l|}{\multirow{2}{*}{DT}} & Normal     & \multirow{2}{*}{99.4\%}   &  99.8     & 99.5      & \textbf{ 0.9965}        \\ \cline{2-2} \cline{4-6} 
\multicolumn{1}{|l|}{}                     & Attack     &                           &  97.9     & 99.1      & \textbf{0.9850}         \\ \hline
\multicolumn{1}{|l|}{\multirow{2}{*}{RF}} & Normal     & \multirow{2}{*}{99.5\%}   &  99.9     & 99.6      &  \textbf{0.9965}        \\ \cline{2-2} \cline{4-6} 
\multicolumn{1}{|l|}{}                     & Attack     &                           &  98.2     & 99.5      &  \textbf{0.9885}        \\ \hline
\multicolumn{1}{|l|}{\multirow{2}{*}{ANN}} & Normal     & \multirow{2}{*}{99.5\%}   &  99.5     & 99.8      &  \textbf{0.9965}       \\ \cline{2-2} \cline{4-6} 
\multicolumn{1}{|l|}{}                     & Attack     &                           &  99.2     & 98.0      &  \textbf{0.9860}        \\ \hline
\end{tabular}
\end{table}

In another experiment, we assessed the performance of ML methods for attack identification, and the results are presented in Table \ref{C:tab:results-multiclass}. The RF method outperformed the others with an accuracy of 98.4\%. However, the ML methods exhibited a decrease in accuracy due to their inability to recognize certain attack types. Although all three methods remained above 98.1\% accuracy, there was a significant decline in F1-score, which is a more reliable assessment metric when dealing with imbalanced classes. Despite this, the ML methods effectively identified normal flows, as evidenced by their F1-score exceeding 0.99. The high F1-score for DDoS attack identification was expected since DDoS flows are easily detectable by monitoring massive packets and connections during the attack. Conversely, the F1-scores for IP-scan, Port-Scan, MitM, and Replay attack indicate that accurately classifying flows into the correct attack types is a non-trivial task, especially for IP-Scan with an F1-score of only 0.52.
\begin{table}[]
\centering
\caption{Test results of three ML-methods for attack identification }
\label{C:tab:results-multiclass}
\begin{tabular}{|ll|l|l|l|l|}
\hline
\multicolumn{2}{|l|}{}                                  & \textbf{Accuracy}                  & \textbf{Precision} & \textbf{Recall}    & F1-\textbf{score} \\ \hline
\multicolumn{1}{|l|}{\multirow{6}{*}{DT}} & Normal     & \multirow{6}{*}{98.1\%}   &  99.8     & 99.5      & \textbf{ 0.9965}        \\ \cline{2-2} \cline{4-6} 
\multicolumn{1}{|l|}{}                     & DDoS       &                           &  100      & 99.9      &  \textbf{0.9995}        \\ \cline{2-2} \cline{4-6} 
\multicolumn{1}{|l|}{}                     & IP-Scan    &                           &  35.9     & 94.8      &   \textbf{0.5208}       \\ \cline{2-2} \cline{4-6} 
\multicolumn{1}{|l|}{}                     & MitM       &                           &  90.9     & 88.2      &  \textbf{0.8953}        \\ \cline{2-2} \cline{4-6} 
\multicolumn{1}{|l|}{}                     & Port-Scan  &                           &  94.9     & 93.2      &  \textbf{0.9404}        \\ \cline{2-2} \cline{4-6} 
\multicolumn{1}{|l|}{}                     & Replay     &                           &  93.6     & 90.5      &  \textbf{ 0.9240}       \\ \hline
\multicolumn{1}{|l|}{\multirow{6}{*}{RF}} & Normal    & \multirow{6}{*}{98.4\%} & 99.9           & 99.5       &  \textbf{0.9970}        \\ \cline{2-2} \cline{4-6} 
\multicolumn{1}{|l|}{}                     & DDoS   &                   & 100          & 99.9       & \textbf{0.9995}         \\ \cline{2-2} \cline{4-6} 
\multicolumn{1}{|l|}{}                     & IP-Scan &                   &  36.8         & 98.4       & \textbf{0.5357}         \\ \cline{2-2} \cline{4-6} 
\multicolumn{1}{|l|}{}                     & MitM      &                   &  91.1         &  90.4      &  \textbf{0.9075}        \\ \cline{2-2} \cline{4-6} 
\multicolumn{1}{|l|}{}                     & Port-Scan    &                   & 94.9          &  93.3      & \textbf{0.9419}         \\ \cline{2-2} \cline{4-6} 
\multicolumn{1}{|l|}{}                     & Replay      &                   &   95.4        &  89.9      & \textbf{0.9257}         \\ \hline
\multicolumn{1}{|l|}{\multirow{6}{*}{ANN}} & Normal    & \multirow{6}{*}{98.2\%} &  99.5         & 99.8       &  \textbf{0.9964}        \\ \cline{2-2} \cline{4-6} 
\multicolumn{1}{|l|}{}                     & DDoS   &                   &   99.9        &  99.7      &  \textbf{0.9980}        \\ \cline{2-2} \cline{4-6} 
\multicolumn{1}{|l|}{}                     & IP-Scan &                   &   36.5        &  65.6      &  \textbf{0.4690}        \\ \cline{2-2} \cline{4-6} 
\multicolumn{1}{|l|}{}                     & MitM      &                   &  90.7         &  92.3      &  \textbf{0.9149}        \\ \cline{2-2} \cline{4-6} 
\multicolumn{1}{|l|}{}                     & Port-Scan    &                   &  93.3         & 89.6       &  \textbf{0.9141}        \\ \cline{2-2} \cline{4-6} 
\multicolumn{1}{|l|}{}                     & Replay      &                   &  98.9         &  88.6      &   \textbf{0.9347}       \\ \hline

\end{tabular}
\end{table}

\begin{figure}[hb!]
\centering
        \includegraphics[width=0.6\linewidth]{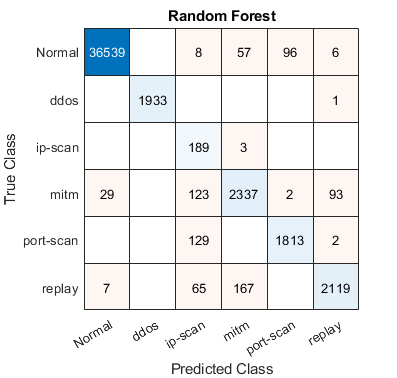}
        \caption{Confusion matrix of RF model for attack identification}
        \label{C:fig:confusion-matrix}
\end{figure}

Figure \ref{C:fig:confusion-matrix} depicts the RF classifier's confusion matrix to analyze misclassification samples. The majority of mistakes are caused by either: 1) attack records that were mistakenly classified as IP-Scan attacks, and 2) confusion between records of replay and MitM attacks. After conducting an in-depth technical analysis of the Port-Scan, Replay, and MitM attacks, it was discovered that they employ the ARP poisoning technique as part of their process, which is remarkably similar to the IP-Scan process. This similarity in process results in the misclassification of some flows. Additionally, both the Replay and MitM attacks follow a same strategy for directing packets to the attacker node. Although the MitM attacker node alters routed packets, while the replay attacker node simply records packets temporarily, both attacks have the same routine for route redirection, causing similar impacts on network traffic.

Finally, the findings in this section show that ML methods can detect cyberattacks with high accuracy. However, determining attack types is not always straightforward. Individual flow analysis cannot differentiate between flows of identical processes used in different attack procedures. Although the impacts of various cyberattacks on network parameters are sometimes extremely similar, assessing the network status before and after individual records can provide a more accurate approximation of the type of prospective attack on the system. To tackle the mentioned problem,  analyzing the predecessor and successor is a promising technique, which is the primary goal of sequence anomaly detection techniques.

\section{Threats to validity}
\label{C:sec:validity}

To enhance the validity of our research against potential challenges, we have adopted the well-established guidelines outlined in \cite{feldt2010validity}. As part of this process, we have identified potential validity threats and implemented appropriate mitigation strategies to address them.

\subsection{Threats to construct validity}
\label{C:subsec:construct_validity}

In order to simulate cyber threats, we had to select a limited number of attacks. We chose to focus on attacks commonly referenced in previous studies, such as those cited in \cite{faramondi2021hardware,morris2014industrial,gomez2019generation}. These attacks are also recognized as common cyberattacks on ICSs by globally-accessible knowledge bases, such as the \textit{'MITRE ATT\&CK'}\footnote{\url{https://attack.mitre.org/techniques/ics/}} and reports from the European Union Agency for Cybersecurity (ENISA) \cite{enisa2018good}. Moreover, while there is no standard approach or tool for implementing cyber attacks, we simulate attacks using common and open-source tools like Scapy and NMap. This decision was made to reduce the possibility of attack simulation errors, which could compromise the construct validity of our results.

Attack implementation in our analysis assumes that the attacker has already gained access to the control zone network. This type of access can be obtained using various methods, including physical access to network equipment, hijacking ICS wireless communications, or exploiting infected ICS components with malware, as described in \cite{dehlaghi2023icssim}. For example, malicious software, such as trojans and viruses, can enable unauthorized operations on the ICS. In addition, attackers may establish backdoors or use remote access software to access the control zone network remotely.

\subsection{Threats to internal validity}
To mitigate any potential bias towards fake results, we took measures to ensure the integrity of our findings. Expressly, we set aside 30\% of the data as a test dataset and allocated 50\% of the data for training the model and 20\% for validation. This rigorous approach allowed us to avoid any potential bias introduced by hyper-parameter tuning, and the results presented in Section \ref{C:sec:results} can be considered reliable.

Given the lack of a realistic distribution of carried attacks in operating ICSs, we deliberately decided not to balance the classes in our generated dataset between normal and under-attack samples. Instead, we utilized various performance metrics such as F1-score, precision, recall, and confusion matrix to assess the efficacy of our ML methods on unbalanced data. This approach allowed us to accurately evaluate the performance of our model without artificially manipulating the data distribution.

\subsection{Threats to internal validity}

We have made significant contributions to the ICSSIM Framework \cite{dehlaghi2023icssim} by enhancing it to create a network dataset that other researchers can utilize to develop IDSs. Our ICS-Flow dataset was created by identifying relevant network properties that could be employed in other ICSs to detect intrusions. To facilitate the creation of this dataset, we developed an open-source ICSFlowGenerator tool that can calculate network flow features from raw network traffic data. These features can be extended or utilized in similar research to convert network traffic into a network flow dataset. Furthermore, the intrusion detection ML models we have described can be retrained with new environment network data for different ICSs. This opens up opportunities for other researchers to build upon our work and adapt it to their use cases.

\section{Conclusion}
\label{C:sec:conclusion}

The ICS-Flow dataset is introduced in this paper as a benchmark for validating ML-based network intrusion detection techniques in ICSs. The dataset was created using ICSSIM Tools to set up a virtual ICS testbed for a sample bottle filling factory. To simulate realistic attacks on ICSs, we employed various common attack types, drawing from observations in `ENISA' and `MITRE ATT\&CK'. During both normal and attack scenarios, we recorded the ICS's network packets and physical process state variables. To handle the computational complexity of analyzing individual network packets, we developed the ICSFlowGenerator tools as a free and open-source solution for processing captured raw network data into a network flow dataset. This tool can analyze network packets and produce network flows that include 50 different network features, such as flow features, general features, and TCP features. We also labeled the network flow records using various strategies to facilitate supervised learning studies and provide a foundation for testing unsupervised approaches. We have made the raw network traffic, the flow dataset, and log attack files publicly available for research in this field. Finally, we evaluated the dataset's applicability for intrusion detection validation using several supervised ML techniques, including ANN, DT, and RF.

Implementing IDS in industrial systems poses a number of challenges, with false alarms being the most significant obstacle. These false alarms can lead to costly interruptions in regular system operations. To mitigate this issue, a promising direction for future research is to explore the use of extended monitoring periods or sequence anomaly detection techniques, which have the potential to reduce the incidence of false alarms. Another challenge associated with IDS implementation is the unpredictability of attacker techniques. Attackers may use novel techniques that cause different effects on network packets, making it impossible to detect them using supervised ML. For this reason, investigating unsupervised binary and multiclass classification is another direction for future work.

Moreover, it is worth noting that network attacks not only impact network traffic but can also modify physical processes. As the ICS-Flow dataset includes both types of data, a potential direction for future research would be to integrate network monitoring with physical process monitoring to identify cyberattacks.

\section*{Acknowledgment}

This work was supported by InSecTT (www.insectt.eu), which received funding from the KDT Joint Undertaking (JU) under grant agreement No 876038. The JU receives support from the European Union’s Horizon 2020 research and innovation programme and Austria, Sweden, Spain, Italy, France, Portugal, Ireland, Finland, Slovenia, Poland, Netherlands, Turkey, Belgium, Germany, Czech Republic, Denmark, Norway. 

The document reflects only the authors’ views and the Commission is not responsible for any use that may be made of the information it contains.

\bibliographystyle{IEEEtran}
\bibliography{IEEEabrv,refs}

\end{document}